\documentclass[10pt, conference]{IEEEtran}
\IEEEoverridecommandlockouts

\usepackage[
  backend=biber,
  style=ieee,
  minbibnames=100, maxbibnames=100,
]{biblatex}
\AtBeginDocument{} % Setzt den richtigen Titel

\usepackage{multirow}

\usepackage{amsmath,amssymb,amsfonts}
\usepackage{graphicx}
\usepackage{subcaption} 
\usepackage{textcomp}
\usepackage{xcolor} 
\usepackage[hidelinks, bookmarks=false]{hyperref}
\usepackage{booktabs,threeparttable}

\usepackage{tikz}
\usepackage{pgfplots}
\pgfplotsset{compat=1.18}
\usetikzlibrary{pgfplots.statistics}
\usetikzlibrary{patterns}
\usepackage{amsfonts}
\usepackage[capitalise]{cleveref}

\usepackage{framed} % Möglicherweise müssen Sie dieses Paket installieren

\usepackage[ruled, lined, linesnumbered, commentsnumbered, longend, algo2e]{algorithm2e}

\usepackage[algo2e,linesnumbered,ruled,lined]{algorithm2e}
\SetKwInOut{Input}{Input}
\SetKwInOut{Output}{Output}
\SetKwComment{Comment}{/* }{ */}

\SetAlFnt{\small\sffamily}

\SetKw{Continue}{continue}
\usepackage{array}
\usepackage{multirow}
\usepackage[table,xcdraw]{xcolor}
\usepackage{stfloats}

\newcolumntype{P}[1]{>{\centering\arraybackslash}p{#1}}

\makeatletter
\def\Cline#1#2{\@Cline#1#2\@nil}
\def\@Cline#1-#2#3\@nil{%
  \omit
  \@multicnt#1%
  \advance\@multispan\m@ne
  \ifnum\@multicnt=\@ne\@firstofone{&\omit}\fi
  \@multicnt#2%
  \advance\@multicnt-#1%
  \advance\@multispan\@ne
  \leaders\hrule\@height#3\hfill
  \cr}
\makeatother

\usepackage[acronym]{glossaries}
\newacronym{dag}{DAG}{Directed Acyclic Graph}
\newacronym{qor}{QoR}{Quality of Result}
\newacronym{qos}{QoS}{Quality of Service}
\newacronym{qoe}{QoE}{Quality of Experience}
\newacronym{yolo}{YOLO}{You Only Look Once}
\newacronym{rcnn}{R-CNN}{Region-based Convolutional Neural Network}
\newacronym{hog}{HOG}{Histograms of Oriented Gradients}
\newacronym{cnn}{CNN}{Convolutional Neural Network}
\newacronym{iot}{IoT}{Internet of Things}
\newacronym{iiot}{IIoT}{Industrial Internet of Things}
\newacronym[plural=FMs,firstplural=Feature Models]{fm}{FM}{Feature Model}
\newacronym[plural=SPLs,firstplural=Software Product Lines]{spl}{SPL}{Software Product Line}
\newacronym{JMH}{JMH}{Java Microbenchmark Harness}
\newacronym{cma}{CMA}{Cumulative Moving Average}
\newacronym[plural=CTCs,firstplural=Cross-Tree Constraints]{ctc}{CTC}{Cross-Tree Constraint}
\newacronym{dspl}{DSPL}{Dynamic Software Product Line}
\newacronym[plural=EFMs,firstplural=Extended Feature Models]{efm}{EFM}{Extended Feature Model}
\newacronym[plural=PFMs,firstplural=Parial Feature Models]{pfm}{PFM}{Partial Feature Model}
\newacronym{cnf}{CNF}{Conjunctive Normal Form}
\newacronym[plural=NFVs,firstplural=Network Functions Virtualizations]{nfv}{NFV}{Network Functions Virtualization}
\newacronym[plural=RQs,firstplural=Research Questions]{rq}{RQ}{Research Question}
\newacronym{v2x}{V2X}{Vehicle-to-Everything}
\newacronym{sat}{SAT}{Boolean Satisfiability Problem}
\newacronym[plural=NPUs,firstplural=Neural Processing Units]{npu}{NPU}{Neural Processing Unit}
\newacronym{RAPL}{RAPL}{Running Average Power Limit}
\newacronym{Kepler}{Kepler}{Kubernetes-based Efficient Power Level Exporter}
\newacronym{ocr}{OCR}{Optical Character Recognition}
\newacronym{ALPR}{ALPR}{Automatic License Plate Recognition}
\newacronym{SVR}{SVR}{Support Vector Regression}
\newacronym{MAE}{MAE}{Mean Absolute Error}
\newacronym{ETL}{ETL}{Extract–Transform–Load}
\newacronym{MAPE}{MAPE}{Mean Absolute Percentage Error}
\newacronym{OHE}{OHE}{one-hot encoder}
\newacronym[plural=SLOs,firstplural=service-level objectives]{SLO}{SLO}{service-level objective}
\newacronym{CRI}{CRI}{Container Runtime Interface}
\newacronym{IQR}{IQR}{interquartile ranges}
\newacronym{mAP}{mAP}{Mean Average Precision}
\newacronym{RMSE}{RMSE}{Root Mean Squared Error}
\newacronym[plural=MLPs,firstplural=Multilayer Perceptrons]{mlp}{MLP}{Multilayer Perceptron}

\definecolor{rowgray}{gray}{0.85} % z.B. 85% Weiß

\usepackage{balance}
\usepackage{scalerel}
\usetikzlibrary{svg.path}

\definecolor{orcidlogocol}{HTML}{A6CE39}
\tikzset{
  orcidlogo/.pic={
    \fill[orcidlogocol] svg{M256,128c0,70.7-57.3,128-128,128C57.3,256,0,198.7,0,128C0,57.3,57.3,0,128,0C198.7,0,256,57.3,256,128z};
    \fill[white] svg{M86.3,186.2H70.9V79.1h15.4v48.4V186.2z}
                 svg{M108.9,79.1h41.6c39.6,0,57,28.3,57,53.6c0,27.5-21.5,53.6-56.8,53.6h-41.8V79.1z M124.3,172.4h24.5c34.9,0,42.9-26.5,42.9-39.7c0-21.5-13.7-39.7-43.7-39.7h-23.7V172.4z}
                 svg{M88.7,56.8c0,5.5-4.5,10.1-10.1,10.1c-5.6,0-10.1-4.6-10.1-10.1c0-5.6,4.5-10.1,10.1-10.1C84.2,46.7,88.7,51.3,88.7,56.8z};
  }
}

\newcommand\orcidicon[1]{\href{https://orcid.org/#1}{\mbox{\scalerel*{
\begin{tikzpicture}[yscale=-1,transform shape]
\pic{orcidlogo};
\end{tikzpicture}
}{|}}}}

\begin{document}

\title{PRISM: Predictive Runtime In-place Scaling and Model Selection for Edge Microservices}

\author{
    \IEEEauthorblockN{
        Uwe Gropengie\ss{}er%
        \IEEEauthorrefmark{1}\orcidicon{0000-0002-1334-8538},
        Thomas Reuter%
        \IEEEauthorrefmark{1},
        Dominik Sch\"on%
        \IEEEauthorrefmark{1}\orcidicon{0000-0003-2704-2852},
        Osama Abboud%
        \IEEEauthorrefmark{2}\orcidicon{0009-0003-0311-1267},
        Xun Xiao%
        \IEEEauthorrefmark{2}\\
        Max \discretionary{M\"uhl-}{h\"auser}{M\"uhlh\"auser}%
        \IEEEauthorrefmark{1}\orcidicon{0000-0003-4713-5327}
    }
    \IEEEauthorblockA{
        \IEEEauthorrefmark{1}
        Technical University of Darmstadt; Darmstadt, Germany\\
        \IEEEauthorrefmark{2}
        Huawei Heisenberg Research Center Munich; Munich, Germany
    }
}

\maketitle
% Seitenzahlen die nächsten zwei Befehle:
\thispagestyle{plain}
\pagestyle{plain}

\begin{abstract}
Latency-sensitive edge AI services must balance strict deadlines, output quality, and limited compute and energy budgets. However, static CPU provisioning wastes resources because inference cost varies substantially across inputs, model variants, and runtime conditions. We present \textsc{PRISM}, a prediction-guided runtime framework that jointly selects model variants and CPU allocations for containerized edge microservices. Using container-level energy monitoring and lightweight regression models, \textsc{PRISM} adapts each pipeline stage in place and minimizes predicted CPU-package energy under deadline, resource, and offline model-level \gls{qor} constraints. We evaluate \textsc{PRISM} on more than $52{,}000$ requests in an \gls{ALPR} pipeline with detection and recognition stages. For detection, \textsc{PRISM} reduces energy consumption by $36\,\%$ compared to the strongest static configuration while preserving a comparable success rate and using less than half of the average CPU allocation. For recognition, it reaches near-static-best performance with lower average CPU allocation. These results show that predictive in-place adaptation is a practical mechanism for making time-sensitive AI microservice pipelines more energy-efficient at the edge.
\end{abstract}


\glsresetall
\section{Introduction}\label{sec:Introduction}
Edge computing relocates computation and storage closer to data sources, reducing latency, lowering bandwidth usage, and enabling real-time analytics~\cite{satyanarayanan_emergence_2017, shi_edge_2016}. At the same time, AI workloads at the edge must operate under tight compute and energy budgets on resource-constrained devices~\cite{wang_deploying_2025}. Many edge inference workloads are latency-sensitive, with \gls{ALPR} in traffic enforcement and access control as a prominent example where late or inaccurate results can hinder incident response or disrupt operations. Consequently, such workloads must respect deadlines, limit energy consumption, and meet application-specific \gls{qor} requirements, which capture task-specific output quality. These requirements differ across domains. Some applications demand near-perfect precision, while others can tolerate approximate results. This trade-off has been widely studied in the context of approximate computing~\cite{liu_retrospective_2020, barua_approximate_2019, chippa_analysis_2013}. In distributed and resource-constrained edge deployments, balancing execution time, energy consumption, and output quality remains difficult because inputs, resource availability, and runtime conditions vary substantially. Our work addresses this setting by making energy--time--quality trade-offs explicit and actionable for containerized edge AI services.

% Plots wurden generiert aus dem Datensatz 2025-06-22
% Es handelt sich um die prepared_data.csv im Subdir energy-model/all-ms-models
% Hintergrund: Dies sind die geglätteten Daten die durch die Python Funktion mscale.energy_model.common.CommonFunctions.load_and_prepare_data geschrieben wurde, es sind die Quelldaten für unsere Energiemodelle
\begin{figure}[tp]
  \centering

  \begin{subfigure}[t]{0.24\textwidth}
    \centering
    \includegraphics[width=\textwidth]{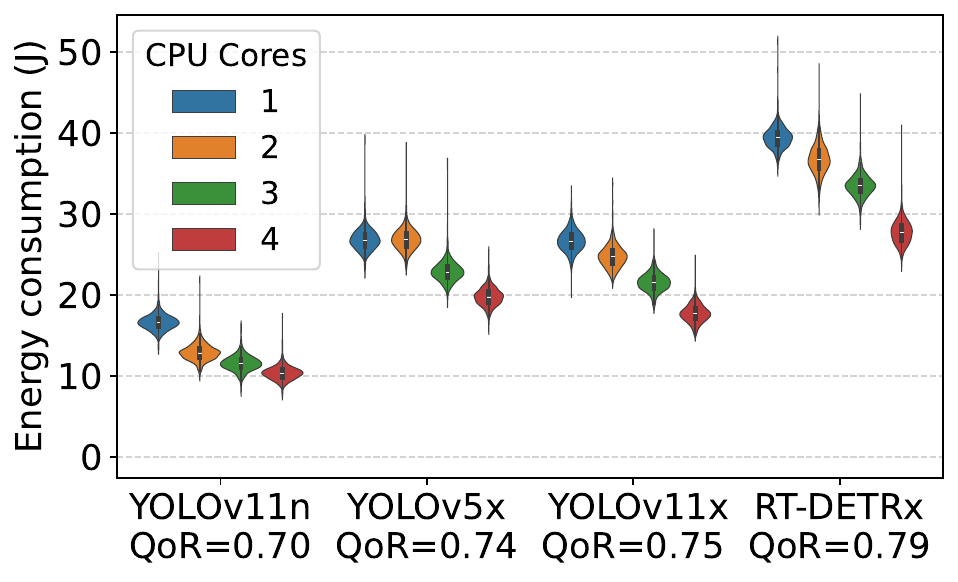}
    \caption{Energy vs CPU cores}
    \label{subfig:energy}
  \end{subfigure}
  \hfill
  \begin{subfigure}[t]{0.24\textwidth}
    \centering
    \includegraphics[width=\textwidth]{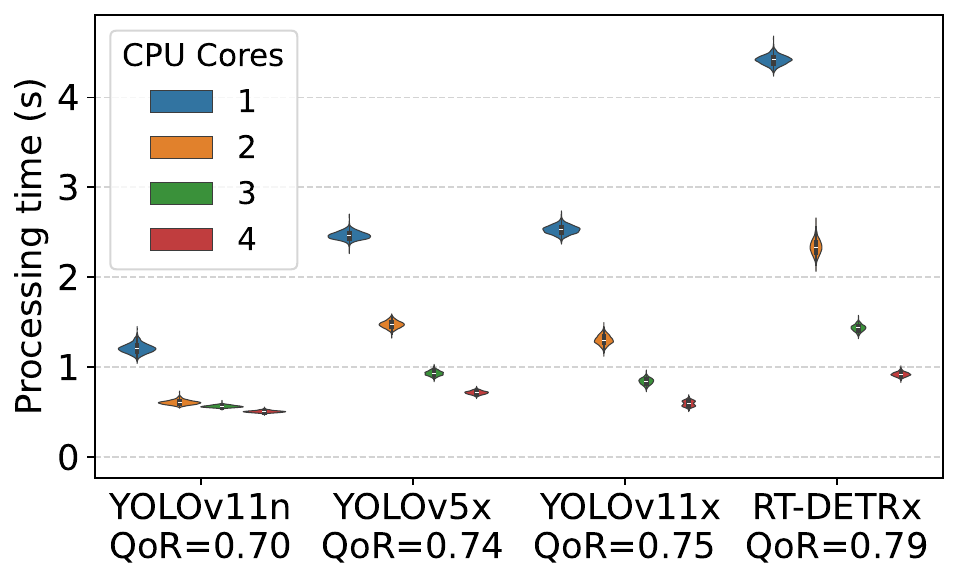}
    \caption{Time vs CPU cores}
    \label{subfig:duration}
  \end{subfigure}
    \caption{Vertical scaling via CPU-core allocations. Distributions of (a) energy and (b) execution time for RT-DETRx, YOLOv11n, YOLOv11x, and YOLOv5x, based on 15{,}118 executions at $6960\times4640$ pixels (32\,MP). The measurement methodology follows prior measurement work~\cite{gropengieser_towards_2026}.}
    \label{fig:EnergyAndTime}
    \vspace{-2mm}
\end{figure}

A first illustration of this trade-off is given in \Cref{fig:EnergyAndTime}, which shows the impact of vertical CPU scaling~\cite{padala_adaptive_2007} on the energy consumption and execution time of different detector variants. The violin plots highlight how model choice and CPU allocation interact. Larger variants such as YOLOv11x, YOLOv5x, and RT-DETRx consume more energy and run longer than smaller variants such as YOLOv11n, even under the same CPU allocation. This reflects the cost of using more accurate models. Vertical CPU scaling, in contrast, primarily reduces execution time within a fixed model. As shown in \Cref{subfig:duration}, additional CPU resources shift execution-time distributions toward shorter runtimes. The corresponding energy distributions in \Cref{subfig:energy} change less uniformly, indicating that faster execution does not automatically translate into proportional energy savings, consistent with prior observations on power--performance trade-offs under vertical scaling~\cite{krzywda_power-performance_2018}. Scaling can therefore reduce latency, while model selection changes both quality and resource demand. Energy-efficient edge inference must consequently consider model selection and CPU allocation jointly, while enforcing explicit deadline and quality constraints.

Prior research has explored both approximation and scaling as mechanisms for resource optimization. MobiDiC~\cite{pandey_mobidic_2016} exploits approximation to trade result quality for reduced execution time in mobile distributed computing, while MobiQoR~\cite{Li2017} introduces \gls{qor} as an optimization dimension for task offloading decisions between local and remote execution. Further work demonstrates the benefits of vertical scaling for meeting \gls{qos} objectives in AI-based applications~\cite{razavi_tale_2024, kannan_grandslam_2019, razavi_sponge_2024}, and the energy footprint of AI models has attracted growing attention~\cite{getzner_accuracy_2023, kurp_green_2008, centofanti_impact_2024}. Most closely related, \textsc{MARQ}~\cite{gropengieser_marq_2025} enables dynamic adaptation of AI service chains at runtime by selecting among alternative execution paths to jointly optimize \gls{qor}, execution time, and energy. However, \textsc{MARQ} and related approaches optimize at the service-chain level and treat the allocation of each microservice stage as fixed during execution. They therefore leave open how a stage should choose among model variants and CPU allocations per request, how these choices jointly affect latency and energy, and how this relationship can be learned from lightweight pre-execution features and exploited at runtime. The combination of prediction-guided model selection and vertical CPU scaling under explicit \gls{qor} and deadline constraints remains insufficiently explored for containerized edge microservices. At the same time, container-level energy monitoring through Kepler and RAPL makes the energy effects of such decisions observable at runtime~\cite{amaral_kepler_2023, khan_rapl_2018}.

In this paper, we present \textsc{PRISM}, a prediction-driven runtime framework for energy-aware adaptation of containerized edge microservices. Building on prior measurement evidence that per-request execution time and energy under vertical CPU scaling are predictable~\cite{gropengieser_towards_2026}, \textsc{PRISM} closes the loop from measurement and prediction to runtime enforcement. It combines regression-based performance modeling, in-place vertical CPU scaling, and per-stage model selection in Kubernetes. A key novelty is the explicit treatment of \gls{qor} as a stage-local runtime constraint rather than optimizing latency or throughput alone. For each request and microservice stage, \textsc{PRISM} selects the configuration that minimizes predicted CPU-package energy while satisfying a deadline, a resource bound, and an offline model-level \gls{qor} threshold. These offline quality scores avoid the need for ground-truth labels during operation, while predictive models trained from container-level measurements estimate execution time and energy before execution. The selected configuration is then applied in place without restarting the pod.

We evaluate \textsc{PRISM} in a realistic \gls{ALPR} pipeline comprising object detection and \gls{ocr} stages. Our experiments show that prediction-guided, stage-local adaptation can substantially reduce energy consumption while preserving success rates comparable to the strongest static configurations. The results demonstrate that combining predictive modeling with in-place scaling and model selection provides a practical mechanism for energy-efficient execution of time-sensitive AI microservice pipelines at the edge.

After introducing the motivating example and background in \cref{sec:Background}, the remainder of this paper is structured as follows.
\begin{itemize}
    \item \Cref{sec:SystemDesign} presents the closed-loop system design, including container-level measurement, offline model training, runtime monitoring, and in-place adaptation of microservice stages.
    \item \Cref{sec:Approach} formalizes the prediction and selection method, including feature construction, predictor deployment, and per-stage constrained selection of model variant and CPU allocation.
    \item \Cref{sec:Evaluation} presents the dataset and research-question-driven evaluation, covering the effects of vertical scaling, prediction accuracy, operational feasibility, and end-to-end behavior on the \gls{ALPR} pipeline.
\end{itemize}
Finally, we discuss insights and limitations in \cref{sec:Discussion}, review related work in \cref{sec:RelatedWork}, and conclude in \cref{sec:conclusion}.
\section{Background and Motivating Example} \label{sec:Background}

\begin{figure*}[t]
    \centering
    \includegraphics[width=0.75\linewidth]{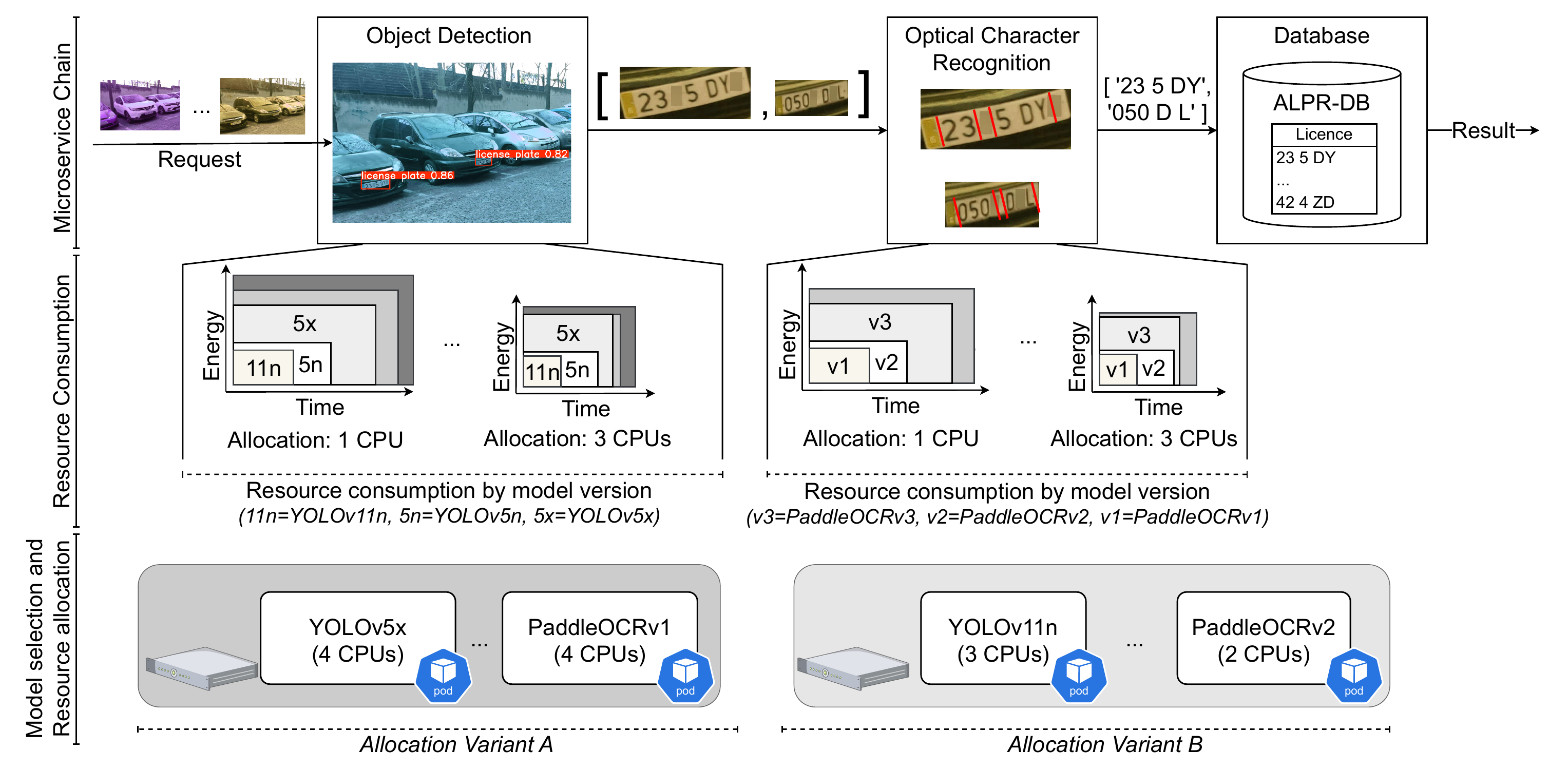}
    \caption{Exemplary presentation of the ``Automated License Plate Recognition'' pipeline. It consists of three stages implemented as microservices. Object detection and OCR support multiple model variants and CPU allocations, while the database stage remains fixed. Different stage assignments and model choices lead to varying energy consumption and execution times.}
    \label{fig:alpr-example}
\end{figure*}

We use \gls{ALPR} as a motivating example to illustrate the challenges of joint model selection and resource allocation in edge-based AI applications. \Cref{fig:alpr-example} shows an \gls{ALPR} pipeline deployed at the edge. We use \emph{stage} to refer to a logical step in the pipeline, such as detection or \gls{ocr}, and \emph{microservice} to describe its containerized implementation. The pipeline consists of three stages implemented as microservices, namely object detection, \gls{ocr}, and a database for license plate validation. We refer to the first two stages as DETECT and OCR. DETECT and OCR offer multiple model variants. For example, DETECT may use YOLOv11n or RT-DETRx, while OCR can rely on lightweight or more advanced recognizers. These choices affect execution time, energy consumption, and \gls{qor}. Smaller models usually reduce latency and energy consumption, whereas larger variants can improve output quality at a higher computational cost.

\textsc{PRISM} treats each stage as an independent adaptation unit. For every request and stage, it selects a model variant and CPU allocation that minimizes predicted CPU-package energy under that stage's deadline, resource bound, and offline model-level \gls{qor} threshold. This stage-local design keeps the runtime decision bounded and predictable and complements application-level orchestration frameworks such as \textsc{MARQ}~\cite{gropengieser_marq_2025}, which operate at the chain level but assume fixed resource provisioning per stage.

While smaller models generally achieve lower execution time and energy consumption, they may fail to provide sufficient quality in accuracy-sensitive applications. In such cases, higher \gls{qor} can justify increased energy use, particularly when latency budgets can still be met through resource scaling. This trade-off underscores the need to jointly consider model selection and resource allocation rather than treating them in isolation. This raises the central question of which variant to select and how to adjust resource allocation to balance output quality, latency, and energy consumption under given constraints.

The runtime selector depends on model-level \gls{qor} constraints and per-request measurements of execution time and CPU-package energy. The following subsections define these signals and explain how they are obtained.

\subsection{Quality of Result}

\gls{qor} is a normalized, percentage-based measure that quantifies the expected output quality of a computation step relative to a task-specific reference metric~\cite{gropengieser_marq_2025, pandey_mobidic_2016}. Its key advantage is that it provides a uniform abstraction for expressing quality requirements across application domains. In computer vision, common metrics include \gls{mAP} for object detection~\cite{padilla_survey_2020} and word-level accuracy for text recognition~\cite{baek_what_2019}. In data analytics, relative error is often used to characterize the quality of approximation~\cite{goiri_approxhadoop_2015, wen_approxiot_2018, hu_approximation_2019}.

The \gls{qor} of a service can in principle be determined either offline or online, depending on the availability of ground truth. In this paper, offline refers to evaluation on annotated datasets during training and benchmarking. Online refers to runtime operation on unlabeled live inputs. Offline assessment yields precise and reproducible quality measurements, but is limited to controlled datasets. During online operation, ground truth is typically unavailable. Runtime \gls{qor} can therefore only be approximated indirectly through proxy signals that correlate with accuracy, such as prediction confidence or temporal consistency in video streams. Although such signals are useful in practice, we do not use them here because their calibration without ground truth is task-specific, deployment-dependent, and difficult to validate across systems.

Instead, we derive per-model offline \gls{qor} scores using \gls{mAP} for DETECT and word-level accuracy for OCR. At runtime, these scores are used as minimum model-level quality constraints during configuration selection. This design enables reproducible and deployment-independent decisions without relying on task-specific confidence calibration or online accuracy estimation. It does not provide a per-request quality guarantee. Rather, it ensures that selected configurations use models whose offline reference quality satisfies the requested threshold.

\subsection{Container-Level Energy Monitoring in Kubernetes}

\begin{figure}[tbp]
    \centering
    \includegraphics[width=0.85\linewidth]{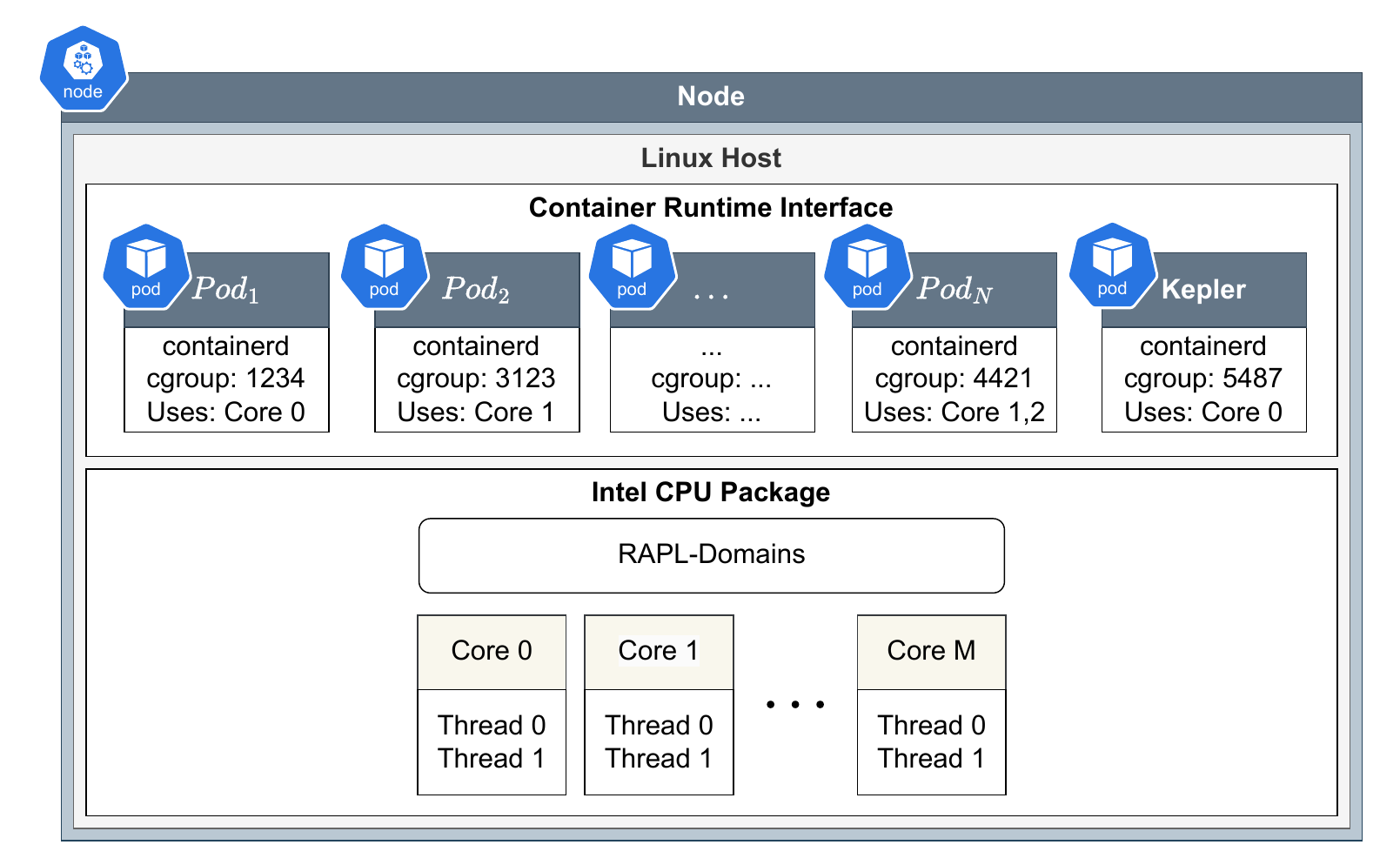}
    \caption{Node-level measurement and attribution path for container energy. Kepler maps CPU-package energy from RAPL counters to pods using per-pod CPU usage, enabling per-request energy accounting in \textsc{PRISM}.}
    \label{fig:nodeArchKepler}
    \vspace{-2mm}
\end{figure}

Precise measurement of execution time and energy is crucial for energy--time--quality-aware adaptation. In this work, vertical scaling refers to adjusting the CPU resources available to a pod and does not cover GPU or TPU resources. We instrument our Kubernetes cluster with Kepler~\cite{amaral_kepler_2023}, which attributes Intel \gls{RAPL}~\cite{khan_rapl_2018} CPU-package energy counters to individual containers. \Cref{fig:nodeArchKepler} illustrates the node-level measurement and attribution path used at runtime. Kepler ingests RAPL package energy and per-pod CPU usage, and attributes package-level energy to pods based on cgroup statistics. \textsc{PRISM} samples these attributed counters immediately before and after each request to derive per-request execution time and CPU-package energy.

Kepler correlates package-level energy from RAPL domains with per-pod CPU usage obtained from Linux cgroups, deriving container-level energy estimates without additional hardware instrumentation. For example, if a pod accounts for 60\,\% of CPU usage in an interval where the CPU package consumed 10\,J, Kepler attributes 6\,J to that pod. This attribution mechanism provides the per-request energy signal on which \textsc{PRISM}'s runtime decisions are based. 

We report CPU-package energy because CPU allocation is the control knob used by \textsc{PRISM} and directly affects execution time, deadline feasibility, and the CPU energy consumed during inference. This focus is also consistent with prior large-scale microservice analysis, which shows that microservice runtime performance is substantially more sensitive to CPU interference than to memory interference~\cite{luo_characterizing_2021}. CPU-package energy, therefore, captures the decision-dependent component most directly affected by \textsc{PRISM}'s in-place scaling decisions. Full-system energy additionally includes components such as DRAM, storage, networking, and static platform baselines. These components remain relevant for full-system accounting, but they are not directly controlled by the CPU allocation decisions studied here. Prior work has demonstrated that RAPL-based CPU energy measurements provide a practical foundation for energy-aware algorithm engineering, while non-CPU domains and container-level power models require platform-specific support or modeling and can be less comparable across systems~\cite{beyer_cpu_2020, choochotkaew_robust_2024}.

Our evaluation, therefore, focuses on CPU-based in-place vertical scaling. In-place scaling of accelerator resources is outside the scope of this work. Kubernetes In-Place Pod Resize currently targets CPU and memory resources, whereas accelerators such as GPUs or TPUs are managed through device-specific mechanisms and do not expose the same generic in-place resizing interface~\cite{kubernetes_resize_nodate}.

\subsection{Model Preloading and In-Place Adaptation}

\textsc{PRISM} assumes that all candidate model variants of a microservice stage are preloaded and kept in memory during operation. This design enables in-place adaptation because model switching does not require loading a new model from storage and can be performed on a per-request basis. Keeping models in memory reduces switching latency and avoids model-loading cold starts during request processing, which is important for deadline-sensitive inference.

The static memory footprint of preloaded model weights is not part of \textsc{PRISM}'s optimization objective. Instead, \textsc{PRISM} optimizes per-request, decision-dependent costs within this operating mode, namely execution time and CPU-package energy. Activation tensors and temporary workspace memory are allocated per inference and released afterwards, while model parameters contribute to the persistent footprint. This assumption is practical when the candidate set fits into memory, but it can limit applicability on highly memory-constrained edge devices. Managing model loading, eviction, placement, or container lifecycle decisions is orthogonal to \textsc{PRISM}'s scope and can be handled by the underlying orchestration layer.
\section{System Design}\label{sec:SystemDesign}

\begin{figure}
    \centering
    \includegraphics[width=0.9\linewidth]{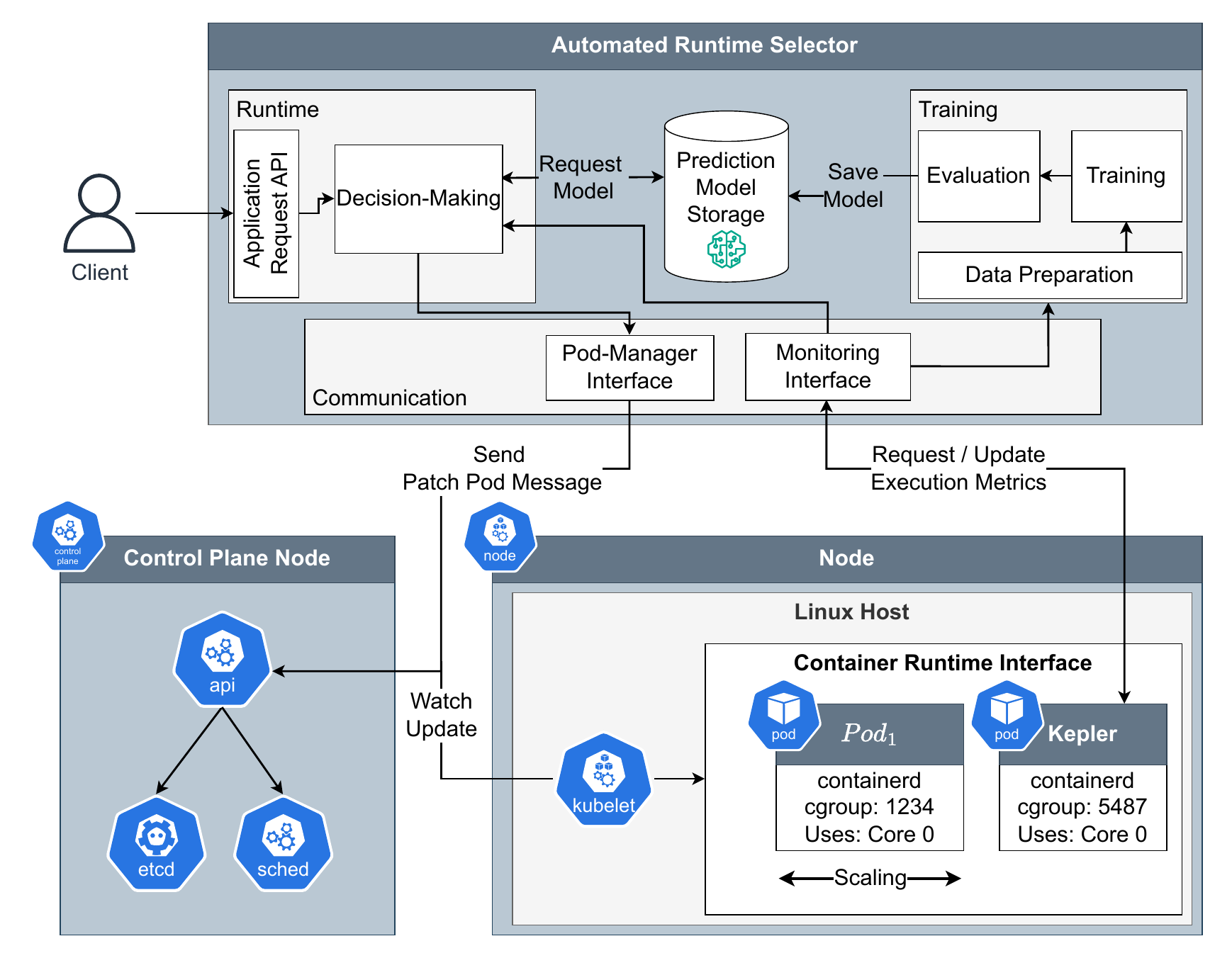}
    \caption{System design of \textsc{PRISM}. The measurement subsystem captures per-request execution time and CPU-package energy via Kepler. Offline preprocessing transforms traces into regression models. At runtime, \textsc{PRISM} filters candidate configurations by offline \gls{qor} and deadline constraints, then applies the energy-minimal selection via in-place CPU scaling and model switching.}
    \label{fig:SystemDesign}
    \vspace{-2mm}
\end{figure}

To enable dynamic, prediction-driven adaptation of microservices at runtime based on execution time, energy consumption, and quality constraints, we design an integrated control loop that adapts containerized edge applications in place. \Cref{fig:SystemDesign} illustrates the overall architecture. The right-hand side depicts the measurement subsystem, which captures fine-grained execution time and energy metrics at the container level on edge hardware. The left-hand side shows the runtime adaptation loop, which evaluates feasible configurations per request and applies the selected one through in-place reconfiguration. Between these two components, an offline preprocessing pipeline transforms collected measurements into predictive models that form the basis for online decision-making. Together, these elements implement control of microservices under realistic operating conditions with awareness of energy, time, and quality. The design deliberately focuses on decision-dependent, per-request costs that can be actuated at runtime. It therefore excludes global pipeline scheduling, model lifecycle management, and system-wide energy components that are not directly controlled by CPU scaling.

The measurement subsystem provides the data foundation for both offline model training and online decision-making. It is deployed as a Kepler DaemonSet on every Kubernetes node and leverages Intel \gls{RAPL} counters to capture CPU package-level energy consumption. These readings are correlated with Linux cgroup statistics and CPU-affinity information obtained via the \gls{CRI}, yielding per-pod time series of execution time and attributed CPU-package energy. While Kepler exposes metrics via Prometheus, the default scrape interval is too coarse for short-lived inference workloads. Therefore, \textsc{PRISM} queries Kepler's metrics directly in Prometheus-exporter format and computes energy differences between consecutive readings at the smallest available interval. This per-request differencing captures millisecond-to-second-scale workloads and, in combination with repeated executions, reduces the impact of transient system noise and outliers. Because energy attribution is derived from CPU-package counters and normalized by per-pod CPU usage, the resulting measurements are aligned with the vertical CPU scaling decisions made by \textsc{PRISM}. This makes the signal more suitable for runtime selection than system-level energy metrics whose static components would reduce sensitivity to CPU allocation changes.

Collected measurements are processed offline in a preprocessing pipeline. Traces are transformed into feature vectors that include the selected model variant, CPU allocation, and input characteristics such as image resolution and file size. Based on these features, regression models predict execution time and CPU-package energy. Result quality is incorporated via precomputed per-model \gls{qor} scores rather than inferred online. This enables reproducible decision-making without task-specific confidence calibration or ground-truth labels at runtime, but constrains quality at the model level rather than guaranteeing per-request quality. Candidate predictors include linear and polynomial regression, \gls{mlp}, XGBoost, and Bayesian regression. Model selection is primarily guided by \gls{MAPE}, which reflects relative prediction error and is therefore relevant for per-request decisions. We additionally report $R^2$, \gls{MAE}, and \gls{RMSE} for completeness. The selected predictors are serialized and stored in a model database together with metadata such as training time and observed accuracy. During operation, predicted and measured values are logged to support later offline retraining in the presence of workload drift.

At runtime, \textsc{PRISM} performs stage-local decision-making for each incoming request. For a given request and stage, all combinations of model variant $m$ and CPU allocation $c$ are enumerated. Candidate configurations are first constrained by offline model-level quality, then pruned by deadline feasibility, and finally ranked by predicted energy consumption. This ordering discards infeasible options early and focuses optimization on decision-dependent costs. For each candidate, the trained regression models predict execution time $\widehat{T}(m,c)$ and energy consumption $\widehat{E}(m,c)$, while quality is checked against the offline per-model score $\mathrm{QoR}_m$. Configurations whose predicted execution time exceeds the request deadline $T^{\max}$ or whose offline quality score falls below the minimum threshold $\mathrm{QoR}^{\min}$ are discarded. Among the remaining feasible options, \textsc{PRISM} selects the configuration with the minimum predicted energy. Decisions are made independently per microservice and per request, without redistributing deadlines or quality budgets across pipeline stages.

The selected configuration is applied \textit{in place} using Kubernetes' In-Place Vertical Scaling mechanism. The kubelet adjusts CPU allocations directly in the container's cgroups without restarting the pod, preserving execution state. Model switching is likewise handled inside the running microservice. Each microservice preloads all supported model weights at initialization, and the active model is selected via an internal switch at request time. After initialization, this design avoids model-loading delays during request processing. The per-request decision overhead is bounded by the deployed model variants, CPU options, and configured solver termination criteria. In our evaluation, even the more expensive XGBoost-Timefold configuration remains below the corresponding inference latency, as shown in \Cref{sec:Evaluation}. By relying on \gls{CRI} calls and in-place reconfiguration, \textsc{PRISM} supports adaptation without pod restarts and keeps runtime behavior consistent across requests.

After execution, the observed execution time and energy consumption are fed back into the measurement infrastructure, closing the control loop. Measurement, offline modeling, runtime prediction, and in-place actuation thus form an integrated system for adapting microservices to varying inputs and constraints on edge hardware. Because the same decision and actuation mechanism is applied independently at each microservice stage, it can be composed along multi-stage pipelines while preserving stage-local decision boundaries.
\section{Approach}\label{sec:Approach}

Building on the system design in \Cref{sec:SystemDesign}, we formalize the predictive modeling and per-stage constrained selection that enable \textsc{PRISM}'s runtime adaptation. The approach addresses three interdependent concerns. First, it defines how pre-execution information is represented as features that capture the joint effect of model variant, CPU allocation, and input properties. Second, it defines how predictors for execution time and energy are trained and deployed. Third, it defines how the per-stage selection problem is solved at request time under deadline, offline quality, and resource constraints.

To predict runtime behavior, we map pre-execution information to features. For each request, we construct a feature vector for each candidate configuration by combining the CPU allocation $c$, the model variant $m$, and observable input properties such as image height $h$, width $w$, and file size $b$. These variables comprise the decision parameters $(m,c)$ and the observed input properties $(h,w,b)$ that influence performance. Together, they define the feature vector
\[
    \mathbf{z}(m,c,h,w,b) = [m,\,c,\,h,\,w,\,b].
\]
\gls{qor} is deliberately not part of the feature vector because request-level ground truth is unavailable during live operation, as discussed in \cref{sec:Background}. Estimating per-request \gls{qor} would require access to ground truth at runtime, which is not feasible in the target edge setting. All features used by \textsc{PRISM} are available before stage execution and can be extracted without running the selected model. This keeps the selector independent of model internals and allows decisions to be made before inference starts.

Instead, we precompute average quality scores $\mathrm{QoR}_m$ for each model on a labeled reference dataset and compare them against per-stage thresholds $\mathrm{QoR}^{\min}_k$ provided by the application or request policy at runtime. This design supports reproducible selection decisions and avoids task-specific online proxies. It constrains the selected model by its offline reference quality rather than guaranteeing the quality of every individual request.

Because the features differ in type and scale, preprocessing is required before training and inference. The categorical model identifier $m$ is converted with one-hot encoding so that each architecture is represented by its own binary dimension, allowing downstream regressors to consume $m$ as numeric input. The numeric variables $(c,h,w,b)$ are standardized to reduce scale effects and improve numerical stability. The standardization parameters are fitted on the training data and reused unchanged at runtime to keep the training and inference distributions consistent.

Polynomial terms are only used when training explicit polynomial regression baselines. They provide linear models with additional flexibility to capture simple nonlinear relationships such as diminishing returns from additional CPU allocation. For a scalar input $u$, we use a polynomial feature map of degree $d$
\[
    \phi_d(u) = (u^1,\ldots,u^d),
\]
which adds powers of $u$ only and does not include interaction terms. Using this notation, the full preprocessing map for the polynomial baselines is
\[
    \varphi(\mathbf{z}) = \bigl[\phi_d(c),\,\phi_d(h),\,
    \phi_d(w),\,\phi_d(b),\,\mathrm{OHE}(m)\bigr].
\]
No constant term is included in $\phi_d(\cdot)$ because the intercept is handled by the model's bias parameter. For predictors such as XGBoost and \glspl{mlp}, we apply the same preprocessing steps except that the polynomial terms are omitted, since these models capture nonlinear relationships through trees or learned hidden layers. For example, with polynomial degree $d=3$ and CPU allocation $c=2$, the expansion yields $\phi_3(c)=(2,4,8)$.

We evaluate a family of regression candidates reflecting different bias--variance trade-offs, including linear models, polynomial regression, Bayesian regression, \glspl{mlp}, and XGBoost. XGBoost consistently yields the most accurate predictions for both execution time and energy in our experiments and is therefore deployed as the default predictor in \textsc{PRISM}. The other regressors are used as baselines in the evaluation and are not queried by the default runtime configuration. The full comparison is reported in \Cref{sec:Evaluation}.

Beyond point accuracy, uncertainty can matter when enforcing strict deadlines. Bayesian regression provides credible intervals that quantify posterior uncertainty around a prediction. For $y \in \{E,T\}$ we assume
\[
    \begin{aligned}
        y \mid X &\sim \mathcal{N}(X\beta+\alpha,\ \sigma^2), \\
        \alpha,\ \beta_k &\sim \mathcal{N}(0,\ \tau^2), \\
        \sigma &\sim \mathrm{HalfNormal}(\gamma),
    \end{aligned}
\]
which yields point predictions $\widehat{E}$ and $\widehat{T}$ together with credible-interval half-widths $\Delta T$. At runtime, deadlines can then be treated conservatively by requiring
\[
    \widehat{T} + \Delta T \le T_{\max}.
\]
When a deterministic predictor such as XGBoost is deployed, there is no posterior by construction. In this case, $\Delta T$ can be replaced by an empirical slack $\delta > 0$ derived from validation residuals, for example, the 95th percentile of $|T-\widehat{T}|$ per configuration.

For each stage $k$, we denote by $\Psi_k$ the deployed predictor that maps the preprocessed input $\tilde{\mathbf{z}}$ to estimates of energy, time, and an optional margin
\[
    \Psi_k(\tilde{\mathbf{z}}) =
    \bigl(\widehat{E}_k,\ \widehat{T}_k,\ \Delta T_k\bigr).
\]
The margin $\Delta T_k$ represents either the half-width of a Bayesian credible interval or an empirical slack derived from validation residuals. In our evaluation, we set $\Delta T_k = 0$ and focus on point predictions. The interface nevertheless supports conservative deadline checks when a margin is enabled.

Training and model selection follow the needs of the runtime decision. All candidate models are tuned using $k$-fold cross-validation, with hyperparameters such as tree depth, learning rate, polynomial degree, and regularization strength varied accordingly. We use \gls{MAPE} as the primary selection metric because it is scale-free across the DETECT and OCR targets. After selection, the final model for each target is retrained on the available data and exported together with all preprocessing metadata. Using the same artifacts at runtime ensures that inference applies the same preprocessing pipeline as training.

\subsection{Pipeline Input Propagation}

In a multi-stage pipeline, the input to a downstream stage depends on the output of an upstream stage. In the \gls{ALPR} pipeline, the OCR stage receives bounding-box crops produced by DETECT rather than the original image. The size of these crops varies with scene content and cannot be known before DETECT has executed.

\textsc{PRISM} addresses this dependency through bucketed size estimation. Before DETECT executes, the OCR decision engine can map the original image resolution to a discrete bucket calibrated on training data. Each bucket stores representative crop dimensions observed during profiling. Once DETECT produces the actual crop, the measured crop dimensions replace the bucket estimate for OCR and any remaining downstream stages. This two-phase approach provides a fallback estimate before upstream outputs are available while using actual dimensions whenever they have been produced.

Bucketing introduces estimation error because the downstream input is only approximated before the upstream stage has completed. This error can be covered by the optional margin $\Delta T_k$ when conservative deadline enforcement is enabled. In our evaluation, we report the point-prediction setting with $\Delta T_k = 0$. Because OCR crops are typically much smaller than the original image, downstream prediction is less dominated by full-resolution image size than the DETECT stage.

\subsection{Decision Overhead}

Requests traverse a microservice chain of $K$ stages. At each stage $k$, the decision engine considers all combinations of model variant $m \in \mathcal{M}_k$ and CPU allocation $c \in \mathcal{C}_k$. The remaining CPU budget $C_{\text{rem}}$ is treated as an external bound supplied by the orchestration layer. \textsc{PRISM} does not decide cluster-wide placement, but uses this bound to avoid selecting configurations that exceed the resources made available to the stage. Given a per-stage deadline $T^{\max}_k$ and a minimum offline quality threshold $\mathrm{QoR}^{\min}_k$, the engine solves
\[
    \begin{aligned}
        \min_{m\in\mathcal{M}_k,\,c\in\mathcal{C}_k} \quad &
        \widehat{E}_k(m,c) \\[2pt]
        \text{s.t.}\quad
        & \widehat{T}_k(m,c) + \Delta T_k(m,c) \le
        T^{\max}_k,\\
        & \mathrm{QoR}_{m} \ge \mathrm{QoR}^{\min}_k,\\
        & c \le C_{\text{rem}}.
    \end{aligned}
\]
The objective minimizes predicted CPU-package energy among all feasible candidates. Execution time appears as a hard deadline, \gls{qor} as an offline model-level quality constraint, and the CPU budget as a strict resource bound. In the \gls{ALPR} pipeline evaluated in \Cref{sec:Evaluation}, the candidate space amounts to $|\mathcal{M}_k| \times |\mathcal{C}_k|$ configurations per stage, with $4 \times 5 = 20$ candidates for DETECT and $4 \times 8 = 32$ for OCR.

\begin{algorithm2e}[t]
\DontPrintSemicolon
\caption{Sequential decision in a microservice chain with offline QoR and optional time margin $\Delta T_k$.}
\label{alg:decision}
\KwIn{Input image $I$ , stages $L$ , per-stage candidates $(M_k, C_k)$ , bounds $(T_k^{\max}, \mathrm{QoR}_k^{\min})$ , remaining CPU budget $C_{\text{rem}}$}
\KwOut{$(m_k,c_k)_k$ or \textsc{fallback}}

Extract $h,w,b$ from $I$\;
\For{$k \in L$ in order}{
  \textit{candidates} $\leftarrow [\,]$\;
  $p_k \leftarrow \texttt{stage\_inputs}(k,\,h,\,w,\,b)$ \tcp*{source-agnostic: raw tuple or derived}
  \For{$m \in M_k$}{
    \uIf{$\mathrm{QoR}_m < \mathrm{QoR}^{\min}_k$}{\Continue}
    \For{$c \in C_k$}{
      \uIf{$c > C_{\text{rem}}$}{\Continue}
      $\tilde{\mathbf{z}}\leftarrow \varphi([c,\,p_k,\,m])$ \tcp*{OHE$(m)$ + standardization; $\phi_d$ only for polynomial baselines}
      $(\widehat{E},\,\widehat{T},\,\Delta T_k)\leftarrow \Psi_k(\tilde{\mathbf{z}})$ \tcp*{$\Delta T_k$: Bayesian half-width or empirical slack}
      \uIf{$\widehat{T}+\Delta T_k > T^{\max}_k$}{\Continue}
      append $(m,c,\widehat{E})$ \textbf{to} \textit{candidates}\;
    }
  }
  \uIf{\textit{candidates} $\neq \emptyset$}{
    $(m_k,c_k) \leftarrow \arg\min_{(m,c,\widehat{E})\in\textit{candidates}} \widehat{E}$\;
    \texttt{apply\_config}($k,(m_k,c_k)$)\;
    \texttt{execute\_stage}($k$)\;
  }\uElse{
    \Return \textsc{fallback}
  }
}
\Return $\{(m_k,c_k)\}_{k\in L}$\;
\end{algorithm2e}

Algorithm~\ref{alg:decision} instantiates this selection as a conceptual enumeration over feasible candidates. In the current deployment, it is realized via Timefold~\cite{de_smet_timefold_2023} with a Hill Climbing local search that terminates early once no further improvement is found within a configurable step limit. The empirical overhead of this implementation is characterized in \Cref{sec:Evaluation}. Because the candidate set is bounded by the deployed model variants and CPU options, direct enumeration is equivalent and preferred for latency-critical deployments.

The loop proceeds stage by stage. It first extracts $(h,w,b)$ from the incoming request and derives the stage-specific inputs $p_k$, either from the raw image tuple or from bucketed downstream sizes when $k$ depends on an upstream output. For each model $m\in \mathcal{M}_k$ satisfying $\mathrm{QoR}_m \ge \mathrm{QoR}^{\min}_k$ and each CPU allocation $c\in \mathcal{C}_k$ with $c \le C_{\mathrm{rem}}$, the loop builds the preprocessed vector $\tilde{\mathbf{z}}=\varphi([c,\,p_k,\,m])$ and queries the deployed predictor via $\Psi_k$ to obtain $(\widehat{E}_k,\widehat{T}_k,\Delta T_k)$. Candidates with $\widehat{T}_k+\Delta T_k > T^{\max}_k$ or violations of the budget or offline quality threshold are discarded. The remaining set is ranked by $\widehat{E}_k$, and the minimizer $(m_k,c_k)$ is selected.

The chosen configuration is enforced in place. The kubelet updates cgroup CPU limits without restarting the pod, and the active model is switched in process because all stage-specific weights are preloaded. After executing stage $k$, the observed outputs are propagated to the next stage and replace the bucketed estimates where applicable. $C_{\mathrm{rem}}$ is provided by the orchestrator and acts as a per-stage resource bound in the selection loop. If no feasible candidate exists, the request is not forwarded to the current stage and is recorded as a failed execution. This case arises when the specified deadline and offline \gls{qor} threshold cannot be satisfied by any available configuration, and is accounted for in the success rate reported in \Cref{sec:Evaluation}.

Finally, measured energy and time are fed back into the monitoring subsystem and compared to the predictions. Deviations exceeding a tolerance threshold can trigger corrective actions such as offline retraining to maintain prediction accuracy under workload drift. This closes the loop of measurement, prediction, optimization, and enforcement.

\section{Evaluation} \label{sec:Evaluation}  

Edge AI workloads must simultaneously satisfy deadlines, energy budgets, and quality constraints under varying inputs and resource configurations. Our evaluation investigates whether predictive models can capture these effects from lightweight pre-execution features and whether \textsc{PRISM} can exploit them for stage-local adaptation in a live microservice chain. This leads to the following research questions.

\textbf{RQ1 -- Prediction Foundation.} How does vertical CPU scaling affect the execution time and energy consumption of edge microservices across model variants and input resolutions, and can these effects be predicted accurately from lightweight pre-execution features?

\textbf{RQ2 -- Operational Feasibility.} What is the per-request decision overhead of \textsc{PRISM} in the deployed microservice chain, and how does cold-start behavior affect latency and energy at pipeline startup?

\textbf{RQ3 -- Pipeline Execution.} How well can \textsc{PRISM} satisfy stage-local deadline and offline \gls{qor} constraints in a realistic pipeline, and how does prediction-guided adaptation compare to static configurations in terms of energy and success rate?

We use the \gls{ALPR} pipeline introduced in \Cref{sec:Background} as the evaluation workload. This structure reflects a common pattern in edge AI applications where detection is followed by recognition or analysis. Similar multi-stage patterns appear in video analytics, autonomous perception, industrial inspection, and medical imaging, where early-stage decisions can influence downstream cost, latency, and quality.

\begin{table}[t]
    \centering
    \caption{Offline \gls{qor} values measured as mAP for DETECT and word-level accuracy for OCR.}
    \label{tab:models_qor_ms1_ms2}
    \setlength{\cmidrulekern}{0pt}
    \setlength{\aboverulesep}{0.4ex}
    \setlength{\belowrulesep}{0.4ex}
    \begin{tabular}{@{} l r @{\quad\quad} l r @{}}
        \toprule
        \multicolumn{2}{c}{\textbf{DETECT (MS1)}} 
        & \multicolumn{2}{c}{\textbf{OCR (MS2)}} \\
        \cmidrule(r){1-2} \cmidrule(l){3-4}
        \textbf{Model} & \textbf{QoR} 
        & \textbf{Model} & \textbf{QoR} \\
        \midrule
        RT-DETR-x   & 0.79 & PaddleOCRv4 & 0.61 \\
        YOLOv11x    & 0.75 & PaddleOCRv3 & 0.49 \\
        YOLOv5x     & 0.74 & PaddleOCRv2 & 0.55 \\
        YOLOv11n    & 0.70 & PaddleOCRv1 & 0.41 \\
        \bottomrule
    \end{tabular}
    \vspace{-2mm}
\end{table}

\subsection{Experimental Setup}

Our experiments use a unified measurement workflow that records result quality, per-request execution time, and container-attributed energy for each combination of input, model variant, and CPU allocation. Offline \gls{qor} is precomputed per model on labeled data and applied at runtime as a model-level quality constraint in the selection process (\Cref{tab:models_qor_ms1_ms2}). Energy is measured as CPU-package joules using Kepler~\cite{amaral_kepler_2023}, with \gls{RAPL} counters attributed to pods and differenced around each request (\Cref{sec:Background}). Each configuration is repeated multiple times, and we report medians and \gls{IQR} to reduce sensitivity to noise.

All experiments run on two identical machines with an Intel Core i9-9940X and 64\,GB RAM under Ubuntu 24.04 LTS. Microservices are containerized and orchestrated using Kubernetes (MicroK8s v1.34.1), and in-place vertical scaling is applied by the kubelet (\Cref{sec:SystemDesign}). Predictive models are trained with scikit-learn~\cite{pedregosa_scikit-learn_2011} and PyMC~\cite{salvatier_probabilistic_2016}, and per-request decision problems are encoded with a Timefold-based solver~\cite{de_smet_timefold_2023}. For deployment in the Java-based microservice chain, the trained preprocessing pipeline is exported in ONNX format~\cite{developers_onnx_2021} and loaded at runtime via ONNX Runtime alongside the serialized XGBoost model. This ensures that the preprocessing transformations applied during training are reproduced at inference time. Per-candidate prediction latency remains in the sub-millisecond range on commodity hardware, as reported in RQ2.

Offline \gls{qor} is measured as \gls{mAP} for DETECT and word-level accuracy for OCR on the UC3M-LP dataset~\cite{ramajo-ballester_dual_2024}, which contains 1,975 vehicle images split into 1,580 training and 395 test images. The resulting model-level scores are shown in \Cref{tab:models_qor_ms1_ms2}. To test behavior beyond the profiling distribution in RQ3, we also evaluate on UFPR-ALPR from Laroca et al.~\cite{laroca_robust_2018}. 

% Input from measurements: 2025-06-22-1127_detect_energy_measurement; generated by GenerateTableForPaper_v3.py
\begin{table*}[ht]
\centering
\caption{Median and IQR of per-request energy (J) and runtime (s) for the DETECT stage across all detector variants, CPU allocations, and input resolutions (RQ1). IQR values are given in parentheses.}
\label{tab:rq1_full}
\begin{tabular}{cccccccccccc}
\toprule
& & \multicolumn{2}{c}{3\,MP} &  \multicolumn{2}{c}{9\,MP} &  \multicolumn{2}{c}{20\,MP} &  \multicolumn{2}{c}{32\,MP} &  \multicolumn{2}{c}{64\,MP} \\
\cmidrule(lr){3-4}\cmidrule(lr){5-6}\cmidrule(lr){7-8}\cmidrule(lr){9-10}\cmidrule(lr){11-12}
Model & C & \~{E}\,(J) & \~{T}\,(s) & \~{E}\,(J) & \~{T}\,(s) & \~{E}\,(J) & \~{T}\,(s) & \~{E}\,(J) & \~{T}\,(s) & \~{E}\,(J) & \~{T}\,(s) \\
\midrule
\multirow{5}{*}{\rotatebox{90}{YOLOv11n}} & 1 & 8.6\,(2.0) & 0.57\,(0.27) & 10.1\,(1.4) & 0.72\,(0.12) & 14.3\,(1.9) & 1.04\,(0.11) & 16.8\,(2.2) & 1.20\,(0.12) & 25.4\,(3.8) & 1.76\,(0.13) \\
 & 2 & 6.1\,(1.2) & 0.14\,(0.04) & 6.9\,(1.5) & 0.21\,(0.07) & 10.2\,(1.7) & 0.40\,(0.05) & 13.1\,(1.9) & 0.60\,(0.05) & 21.8\,(3.7) & 1.12\,(0.06) \\
 & 3 & 4.5\,(0.8) & 0.07\,(0.02) & 5.3\,(1.0) & 0.14\,(0.03) & 9.0\,(1.5) & 0.36\,(0.03) & 11.7\,(1.8) & 0.55\,(0.03) & 20.2\,(3.7) & 1.06\,(0.04) \\
 & 4 & 4.3\,(0.7) & 0.07\,(0.02) & 4.9\,(0.7) & 0.12\,(0.02) & 8.0\,(1.3) & 0.31\,(0.03) & 10.5\,(1.5) & 0.50\,(0.03) & 19.2\,(3.8) & 1.03\,(0.04) \\
 & 5 & 4.2\,(0.7) & 0.07\,(0.02) & 4.9\,(0.7) & 0.12\,(0.02) & 8.0\,(1.4) & 0.31\,(0.03) & 10.6\,(1.6) & 0.50\,(0.03) & 19.1\,(3.8) & 1.02\,(0.04) \\
\cmidrule(lr){1-12}
\multirow{5}{*}{\rotatebox{90}{YOLOv11x}} & 1 & 20.5\,(2.7) & 2.12\,(0.21) & 19.1\,(1.9) & 1.84\,(0.15) & 25.2\,(2.8) & 2.50\,(0.12) & 26.9\,(3.2) & 2.53\,(0.12) & 36.4\,(5.2) & 3.24\,(0.13) \\
 & 2 & 18.9\,(2.9) & 1.04\,(0.15) & 16.5\,(1.9) & 0.80\,(0.07) & 23.0\,(2.6) & 1.21\,(0.11) & 24.8\,(2.8) & 1.29\,(0.11) & 34.7\,(5.3) & 1.88\,(0.12) \\
 & 3 & 16.0\,(1.8) & 0.54\,(0.09) & 14.0\,(1.7) & 0.43\,(0.08) & 20.3\,(2.6) & 0.73\,(0.08) & 21.8\,(2.8) & 0.84\,(0.08) & 31.9\,(4.8) & 1.38\,(0.09) \\
 & 4 & 12.4\,(1.6) & 0.26\,(0.06) & 10.9\,(1.6) & 0.24\,(0.05) & 15.9\,(2.1) & 0.45\,(0.05) & 17.9\,(2.4) & 0.59\,(0.06) & 28.0\,(4.2) & 1.11\,(0.06) \\
 & 5 & 12.3\,(1.9) & 0.22\,(0.05) & 10.7\,(1.4) & 0.23\,(0.03) & 16.5\,(2.5) & 0.45\,(0.05) & 18.5\,(2.8) & 0.62\,(0.04) & 28.6\,(5.2) & 1.15\,(0.06) \\
\cmidrule(lr){1-12}
\multirow{5}{*}{\rotatebox{90}{YOLOv5x}} & 1 & 19.7\,(2.4) & 2.02\,(0.11) & 18.4\,(1.9) & 1.80\,(0.12) & 25.2\,(2.8) & 2.41\,(0.11) & 27.0\,(3.1) & 2.46\,(0.11) & 36.6\,(4.9) & 3.12\,(0.12) \\
 & 2 & 19.8\,(3.2) & 1.10\,(0.17) & 18.4\,(1.9) & 0.94\,(0.10) & 25.2\,(2.8) & 1.36\,(0.11) & 27.0\,(3.0) & 1.46\,(0.11) & 37.3\,(4.8) & 2.08\,(0.11) \\
 & 3 & 15.5\,(2.0) & 0.49\,(0.09) & 14.3\,(1.8) & 0.46\,(0.07) & 21.1\,(2.5) & 0.78\,(0.09) & 23.1\,(2.7) & 0.92\,(0.08) & 33.3\,(4.8) & 1.48\,(0.09) \\
 & 4 & 12.7\,(1.8) & 0.28\,(0.05) & 12.2\,(1.8) & 0.30\,(0.06) & 17.6\,(2.3) & 0.53\,(0.05) & 20.0\,(2.6) & 0.71\,(0.06) & 29.9\,(4.3) & 1.25\,(0.06) \\
 & 5 & 12.9\,(1.9) & 0.24\,(0.06) & 11.7\,(1.6) & 0.26\,(0.03) & 17.6\,(2.6) & 0.49\,(0.05) & 19.6\,(2.9) & 0.65\,(0.05) & 29.5\,(4.7) & 1.20\,(0.06) \\
\cmidrule(lr){1-12}
\multirow{5}{*}{\rotatebox{90}{RT-DETRx}} & 1 & 31.4\,(2.9) & 3.79\,(0.41) & 33.9\,(2.1) & 4.04\,(0.16) & 36.9\,(2.5) & 4.24\,(0.15) & 39.5\,(2.8) & 4.42\,(0.14) & 48.5\,(3.7) & 4.94\,(0.15) \\
 & 2 & 29.1\,(4.5) & 1.93\,(0.26) & 30.0\,(4.7) & 1.94\,(0.27) & 33.8\,(4.6) & 2.09\,(0.31) & 36.7\,(4.9) & 2.28\,(0.31) & 46.6\,(8.5) & 2.80\,(0.31) \\
 & 3 & 24.1\,(2.9) & 0.98\,(0.13) & 26.2\,(2.7) & 1.05\,(0.10) & 30.5\,(3.1) & 1.24\,(0.11) & 33.7\,(3.1) & 1.42\,(0.11) & 43.1\,(6.2) & 1.93\,(0.11) \\
 & 4 & 19.8\,(2.1) & 0.49\,(0.08) & 21.4\,(2.3) & 0.55\,(0.07) & 25.1\,(2.7) & 0.73\,(0.07) & 28.0\,(3.2) & 0.91\,(0.07) & 38.0\,(5.3) & 1.42\,(0.08) \\
 & 5 & 19.0\,(2.3) & 0.38\,(0.05) & 20.9\,(2.4) & 0.45\,(0.05) & 24.2\,(2.8) & 0.62\,(0.05) & 27.3\,(3.3) & 0.80\,(0.05) & 37.2\,(5.1) & 1.31\,(0.07) \\
\bottomrule
\end{tabular}
\vspace{-2mm}
\end{table*}

\subsection{RQ1 -- Prediction Foundation}

\Cref{tab:rq1_full}  reports median and IQR of per-request energy and execution time for the DETECT stage across four model variants, five CPU allocations, and five input resolutions from 3\,MP to 64\,MP. Execution time decreases substantially with additional CPU allocation, but the speedup depends strongly on input size and model architecture. At 3\,MP, the speedup from one to five CPUs ranges from $8\times$ for YOLOv11n to $10\times$ for RT-DETRx, while several 64\,MP configurations show diminishing returns beyond three CPUs. Energy follows a less uniform trend. It often decreases with runtime, but the reduction is smaller at high resolutions and can be non-monotonic at intermediate CPU allocations. This confirms that neither execution time nor energy can be explained by CPU allocation alone.

The IQR values show that additional CPU allocation also reduces runtime variability, which is relevant for deadline enforcement. In many configurations, the IQR of execution time shrinks markedly from one to three CPUs. Overall, the measurements show that the best configuration is input- and model-dependent, motivating prediction-guided selection rather than static provisioning.

We next evaluate whether these effects can be predicted from lightweight pre-execution features. All regressors introduced in \Cref{sec:Approach} are evaluated under identical preprocessing and training conditions. Results are computed on a held-out 10\,\% test split, while the remaining 90\,\% are used for five-fold cross-validation. \Cref{tab:rq1_prediction} summarizes the prediction results. XGBoost achieves the lowest relative error across both microservices and targets, with \gls{MAPE} of 2.6\,\% for time and 4.7\,\% for energy on DETECT, and 6.4\,\% for time and 5.5\,\% for energy on OCR. Linear, Ridge, and Lasso underfit, while polynomial regression and the MLP perform competitively for execution time, but degrade for OCR energy. Bayesian regression provides an uncertainty-aware baseline, but its low point accuracy leads to poor selection performance in RQ3.

\textbf{Interpretation.} Vertical CPU scaling reduces execution time and often also energy, but the effect depends on input resolution and model architecture. These effects are predictable from lightweight features when nonlinear interactions are captured. XGBoost provides the most stable accuracy and is therefore used as the default predictor in \textsc{PRISM}.

%DETECT BAYESIAN: 2025_09_24(rank_2_energy1_time1); OCR: 2025_09_25 (rank_3_energy1_time1)

\begin{table*}[b]
\vspace{-2mm}
\centering
\caption{Prediction accuracy for per-request execution time (T) and energy (E) across regression candidates for DETECT and OCR. \gls{MAPE} is the primary decision-relevant metric. Best values within 0.001 of the optimum are bold.}
\label{tab:rq1_prediction}
\resizebox{\textwidth}{!}{%
\begin{tabular}{lrrrrr|rrrrr|rrrrr|rrrrr}
\toprule
& \multicolumn{5}{c|}{\textbf{DETECT} $T$} & \multicolumn{5}{c|}{\textbf{DETECT} $E$} & \multicolumn{5}{c|}{\textbf{OCR} $T$} & \multicolumn{5}{c}{\textbf{OCR} $E$} \\
\cmidrule(lr){2-6} \cmidrule(lr){7-11} \cmidrule(lr){12-16} \cmidrule(lr){17-21}
Model & $R^2$ & MAE & MSE & RMSE & MAPE & $R^2$ & MAE & MSE & RMSE & MAPE & $R^2$ & MAE & MSE & RMSE & MAPE & $R^2$ & MAE & MSE & RMSE & MAPE \\
\midrule
Bayes & 0.676 & 0.437 & 0.301 & 0.548 & 64.189 & 0.777 & 3.355 & 18.658 & 4.320 & 20.497 & 0.207 & 0.632 & 1.123 & 1.060 & 98.257 & 0.668 & 1.497 & 3.396 & 1.843 & 37.072 \\
Lasso & 0.000 & 0.703 & 0.930 & 0.964 & 107.277 & 0.599 & 4.861 & 33.521 & 5.790 & 32.670 & 0.000 & 0.754 & 1.416 & 1.190 & 166.105 & 0.346 & 1.996 & 6.700 & 2.588 & 53.989 \\
Linear & 0.852 & 0.246 & 0.138 & 0.371 & 36.312 & 0.958 & 1.450 & 3.508 & 1.873 & 9.130 & 0.801 & 0.321 & 0.282 & 0.531 & 73.946 & 0.871 & 0.860 & 1.321 & 1.149 & 22.221 \\
MLP & \textbf{0.997} & 0.032 & \textbf{0.003} & 0.055 & 4.010 & 0.981 & 0.889 & 1.555 & 1.247 & 4.847 & 0.974 & 0.100 & 0.037 & 0.192 & 13.499 & 0.937 & 0.570 & 0.642 & 0.801 & 11.638 \\
Poly-2 & \textbf{0.997} & 0.033 & \textbf{0.003} & 0.056 & 4.075 & 0.980 & 0.927 & 1.635 & 1.279 & 5.220 & 0.969 & 0.111 & 0.044 & 0.210 & 18.047 & 0.931 & 0.605 & 0.706 & 0.841 & 12.401 \\
Poly-3 & \textbf{0.997} & 0.027 & \textbf{0.002} & 0.049 & 3.087 & 0.982 & 0.880 & 1.528 & 1.236 & 4.800 & 0.972 & 0.098 & 0.039 & 0.198 & 13.433 & 0.936 & 0.580 & 0.660 & 0.812 & 12.195 \\
Ridge & 0.852 & 0.247 & 0.138 & 0.371 & 36.408 & 0.958 & 1.450 & 3.509 & 1.873 & 9.130 & 0.801 & 0.321 & 0.282 & 0.531 & 73.784 & 0.871 & 0.860 & 1.321 & 1.149 & 22.152 \\
XGB & \textbf{0.998} & \textbf{0.025} & \textbf{0.002} & \textbf{0.047} & \textbf{2.607} & \textbf{0.983} & \textbf{0.870} & \textbf{1.463} & \textbf{1.210} & \textbf{4.730} & \textbf{0.987} & \textbf{0.056} & \textbf{0.018} & \textbf{0.134} & \textbf{6.377} & \textbf{0.984} & \textbf{0.279} & \textbf{0.166} & \textbf{0.408} & \textbf{5.473} \\
\bottomrule
\end{tabular}%
}
\end{table*}

\subsection{RQ2 -- Operational Feasibility}

\begin{figure*}[tbp]
    \centering
    \begin{subfigure}[t]{0.32\textwidth}
        \centering
        \includegraphics[height=3.5cm]{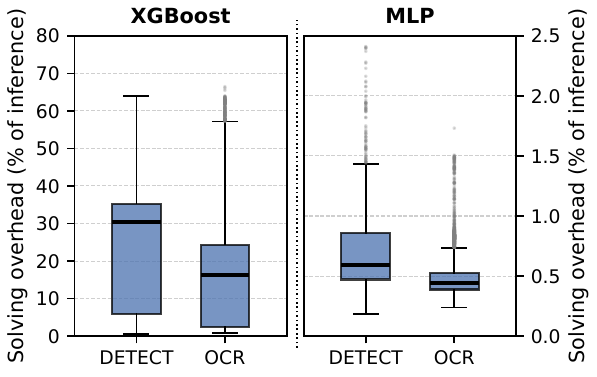}
        \caption{Decision overhead as percentage of inference latency for XGBoost and MLP across DETECT and OCR.}
        \label{fig:rq2_overhead}
    \end{subfigure}
    \hfill
    \begin{subfigure}[t]{0.32\textwidth}
        \centering
        \includegraphics[height=3.5cm]{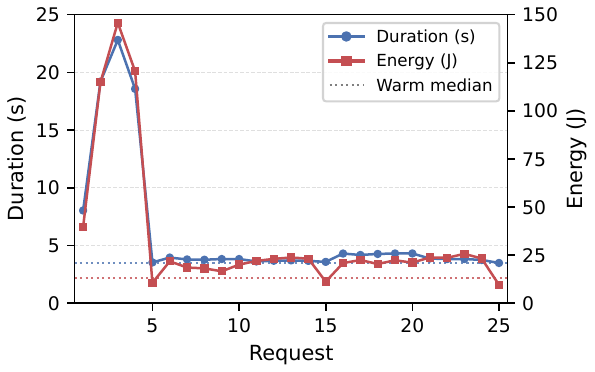}
        \caption{Cold-start effect on duration and energy for DETECT over the first 25 requests. Dotted lines indicate the warm median.}
        \label{fig:rq2_coldstart_detect}
    \end{subfigure}
    \hfill
    \begin{subfigure}[t]{0.32\textwidth}
        \centering
        \includegraphics[height=3.5cm]{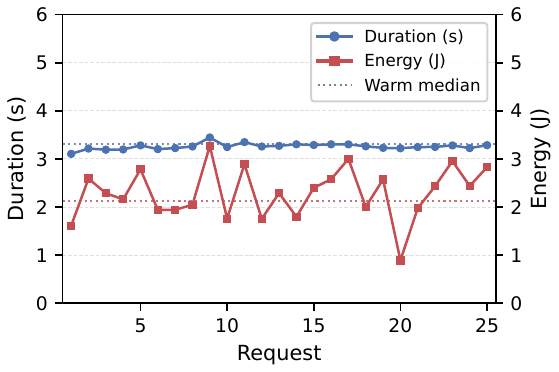}
        \caption{Cold-start effect on duration and energy for OCR over the first 25 requests. Dotted lines indicate the warm median.}
        \label{fig:rq2_coldstart_ocr}
    \end{subfigure}
    \caption{Operational feasibility of \textsc{PRISM} measured across 13,528 DETECT and 9,935 OCR requests.}
    \label{fig:rq2}
\end{figure*}

\Cref{fig:rq2} reports the per-request decision overhead and cold-start behavior of \textsc{PRISM} measured directly in the deployed Java-based microservice chain across 13,528 DETECT and 9,935 OCR requests.

\textbf{Decision overhead.} We measure the overhead of the Timefold-based implementation described in \Cref{sec:Approach}. \Cref{fig:rq2_overhead} reports solving overhead as a percentage of the corresponding inference latency. With XGBoost as the deployed predictor, the median overhead is 30.5\,\% for DETECT and 16.2\,\% for OCR, with 95th percentiles of 59.2\,\% and 42.4\,\%, respectively. This overhead remains below the corresponding inference latency in our deployment, but is large enough to motivate implementation-level optimization. Because the candidate set is bounded by the deployed model variants and CPU options, direct enumeration would be equivalent for this workload and would remove most solver overhead.

With MLP as the predictor, the median overhead drops to 0.6\,\% for DETECT and 0.4\,\% for OCR. However, this reduction comes with lower prediction accuracy and weaker selection quality, as shown in RQ1 and RQ3. The results, therefore, indicate a practical trade-off between predictor quality and decision latency in the current Timefold-based implementation. XGBoost yields more accurate predictions but incurs higher solving overhead, while MLP terminates faster at the cost of solution quality. Both variants operate within a bounded overhead determined by the configured termination parameters.

\textbf{Cold-start behavior.} \Cref{fig:rq2_coldstart_detect,fig:rq2_coldstart_ocr} show the first 25 requests after pod initialization for DETECT and OCR. DETECT exhibits a visible startup effect. The first request takes 8.03\,s and consumes 39.7\,J, corresponding to $2.3\times$ the warm median duration and $3.0\times$ the warm median energy, before both metrics stabilize by request five. OCR shows negligible cold-start behavior, with the first request at $0.9\times$ the warm median duration and $0.8\times$ the warm median energy. This difference reflects the smaller OCR model footprint and shows that startup behavior is stage-dependent. By preloading candidate model weights at initialization as described in \Cref{sec:Background}, \textsc{PRISM} avoids model-loading delays during later request processing.

\textbf{Interpretation.} The decision overhead of \textsc{PRISM} is dominated by the Timefold solver rather than predictor inference cost, as the deployed system records service, model-inference, and solving times separately. XGBoost incurs higher overhead than MLP but yields more reliable configuration selections, confirming the trade-off between prediction quality and decision latency identified in RQ1. Cold-start effects are stage-dependent. They are visible for DETECT but negligible for OCR, indicating that preloading is useful for stages with costly model loading but may be unnecessary for lightweight stages. Startup behavior should therefore be measured per stage and separated from later request processing when assessing deadline feasibility.

\subsection{RQ3 -- Pipeline Execution}

We deploy the full \gls{ALPR} chain with per-request selection of $(m,c)$ at each stage. Decisions remain stage-local and independent across the pipeline. DETECT and OCR models are profiled on UC3M-LP, while the pipeline evaluation uses a mixed Dataset~3 constructed from UC3M-LP and UFPR-ALPR with diverse resolutions and file sizes. Requests are processed with minimum offline \gls{qor} thresholds ranging from 60--79\,\% for DETECT and 40--59\,\% for OCR. Deadlines are chosen from $\{0.25, 0.5, \dots, 3.25\}$\,s. Static baselines execute all fixed $(m,c)$ combinations across the same request set.

All methods are evaluated on $42{,}000$ DETECT and $10{,}500$ OCR requests. Since OCR receives bounding-box crops from DETECT, the effective input size varies per image and detected plate region. Deadlines and offline \gls{qor} constraints are applied independently per stage to isolate stage-local adaptation without global budget redistribution. A request is successful if the selected configuration satisfies the stage-local deadline and offline model-level \gls{qor} constraint. The request set also includes infeasible combinations of deadlines and \gls{qor} thresholds, so even the strongest configuration cannot reach $100\,\%$ success. Per-request energy is taken from Kepler's per-call package-joules and includes failed runs.

\begin{table*}[b]
\caption{Summary of \textsc{PRISM} dynamic predictors on Dataset~3 (42,000 DETECT and 10,500 OCR requests). Energy is total per-request energy including failed executions. Reference: RT-DETRx (5\,CPU) (DETECT), PaddleOCRv4 (1.5\,CPU) (OCR).}
\label{tab:rq3_summary}
\centering
\begin{tabular}{lrrrrrr}
\toprule
    & \multicolumn{3}{c}{DETECT} & \multicolumn{3}{c}{OCR} \\
    \cmidrule(lr){2-4} \cmidrule(lr){5-7}
    Method & Succ.\,(\%) & Avg.\,Cores & Energy\,(kJ) & Succ.\,(\%) & Avg.\,Cores & Energy\,(kJ) \\
\midrule
    XGBoost & 85.7 & 2.4 & 551.1 & 91.0 & 0.6 & 21.5 \\
    MLP & 85.8 & 2.3 & 580.0 & 85.5 & 0.5 & 24.4 \\
    Poly $d{=}2$ & 88.0 & 2.2 & 587.3 & 77.6 & 0.5 & 26.5 \\
    Poly $d{=}3$ & 50.7 & 1.2 & 511.4 & 39.3 & 0.3 & 32.1 \\
    Linear & 80.1 & 2.7 & 558.2 & 73.4 & 0.4 & 17.4 \\
\midrule
    \multicolumn{7}{l}{\textit{Reference: RT-DETRx (5\,CPU) (DETECT), PaddleOCRv4 (1.5\,CPU) (OCR)}} \\
    Static & 84.5 & 5.0 & 859.9 & 90.2 & 1.5 & 21.8 \\
\bottomrule
\end{tabular}
\end{table*}

\begin{figure*}[tbp]
    \centering
    \begin{subfigure}[t]{0.49\textwidth}
        \centering
        \includegraphics[width=\linewidth, height=5cm, keepaspectratio=false]{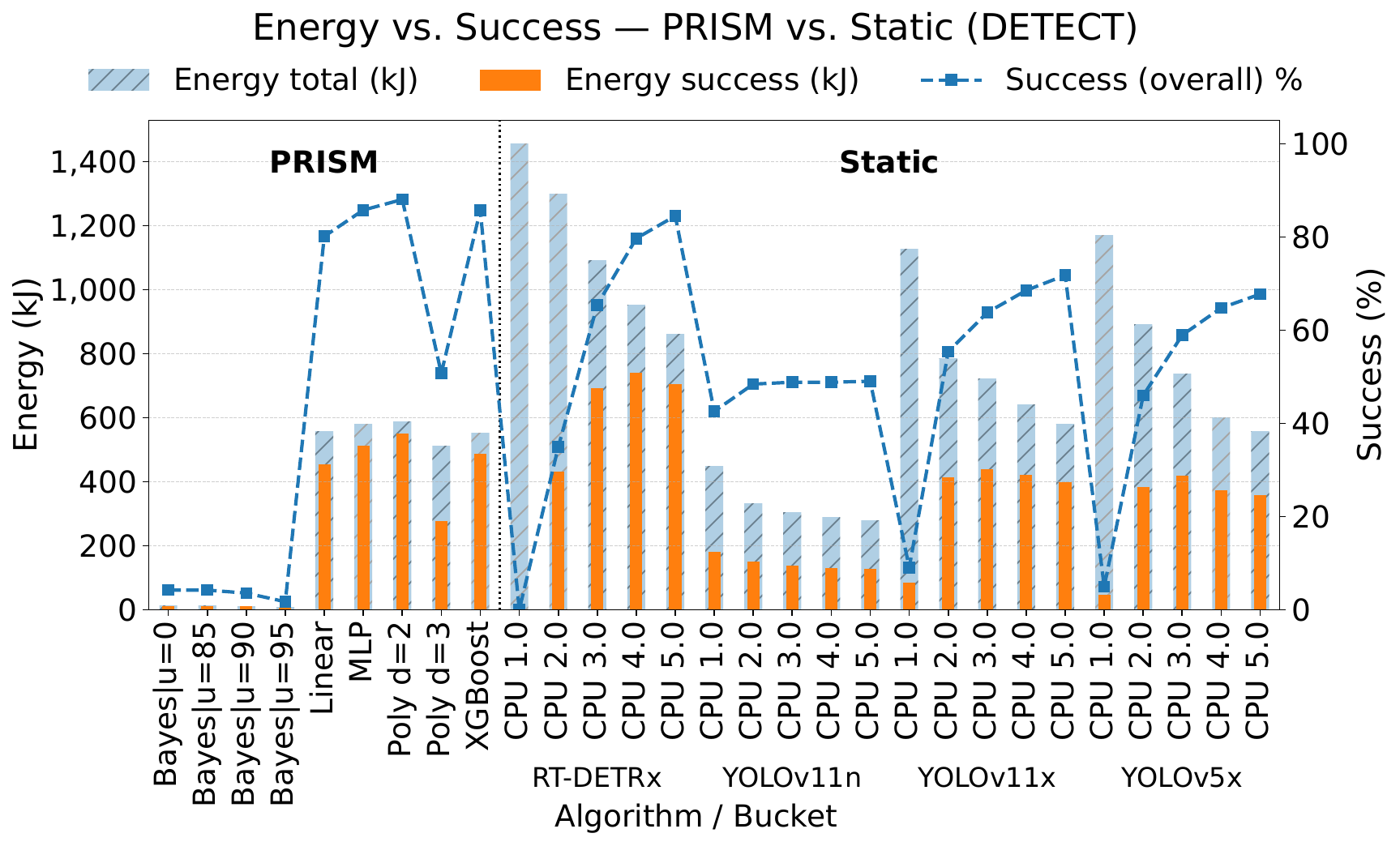}
        \caption{DETECT stage. \textsc{PRISM}+XGBoost matches the best static success rate while reducing total energy by approximately 36\,\% with lower average CPU demand.}
        \label{fig:alpr-eval-detect}
    \end{subfigure}
    \hfill
    \begin{subfigure}[t]{0.49\textwidth}
        \centering
        \includegraphics[width=\linewidth, height=5.1cm, keepaspectratio=false]{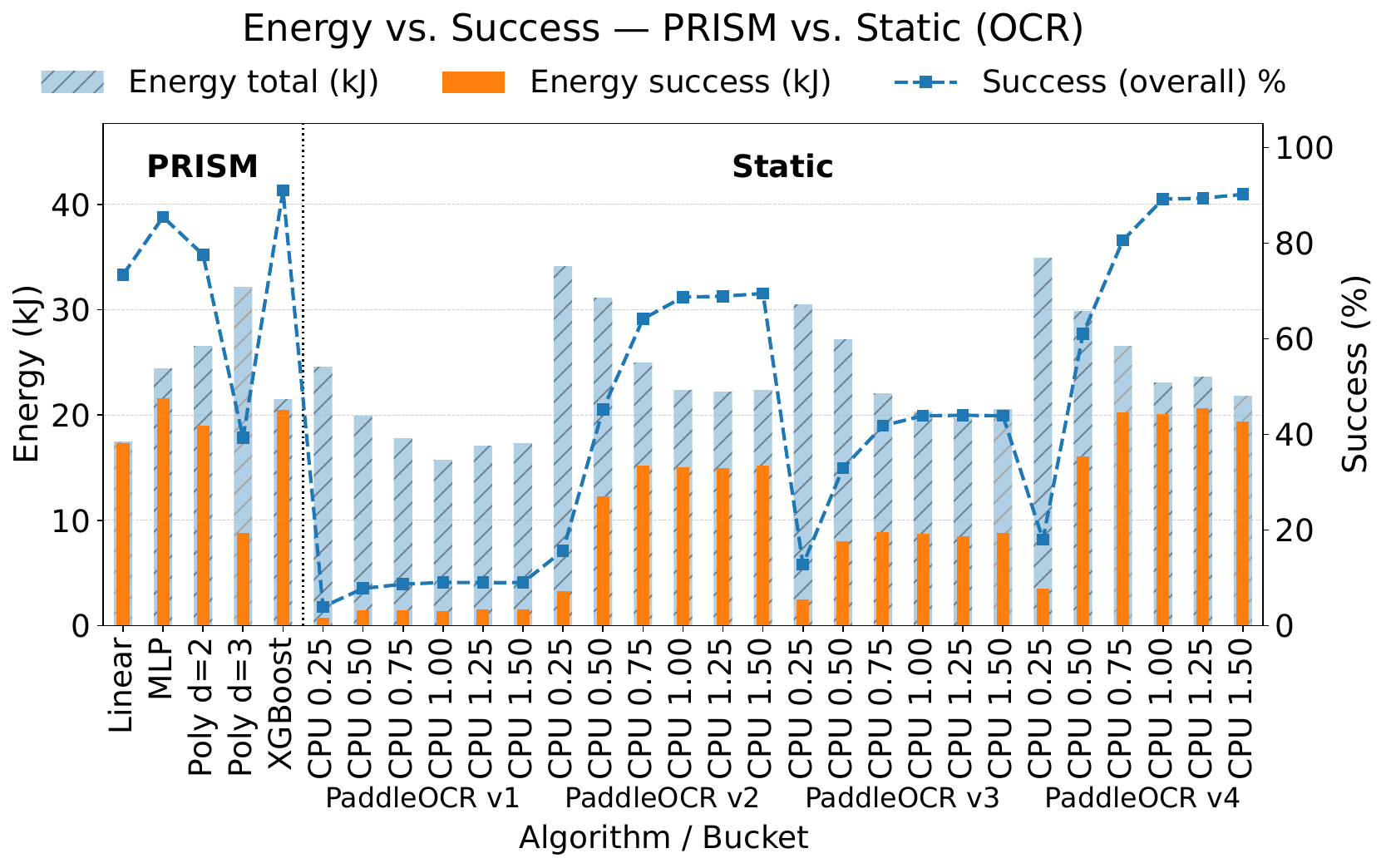}
        \caption{OCR stage. \textsc{PRISM}+XGBoost achieves the best energy--success trade-off among dynamic predictors, with smaller gains than for DETECT because strong static OCR configurations are already close to optimal.}
        \label{fig:alpr-eval-ocr}
    \end{subfigure}
    \caption{Static versus dynamic execution on Dataset~3 for both pipeline stages. Prediction-guided, stage-local adaptation reduces energy consumption while maintaining success rates comparable to the strongest static configurations.}
    \label{fig:alpr-eval}
\end{figure*}

\Cref{fig:alpr-eval} compares dynamic \textsc{PRISM} predictors with static configurations. Bars show total energy and energy spent on successful requests, while the line shows the overall success rate. The left part of each plot contains dynamic predictors, and the right part contains fixed model--CPU combinations. This layout makes the central trade-off visible. Static configurations either use little energy but miss many requests due to deadline or offline \gls{qor} violations, or achieve high success by using expensive models and high CPU allocations. Dynamic selection aims to stay near the high-success region while avoiding unnecessary energy use.

For DETECT, \textsc{PRISM}+XGBoost reaches $85.7\,\%$ success at $551.1\,\mathrm{kJ}$ with an average allocation of $2.4$ CPUs. The strongest static baseline, RT-DETRx with five CPUs, reaches $84.5\,\%$ success at $859.9\,\mathrm{kJ}$. This corresponds to a $35.9\,\%$ energy reduction while preserving a comparable success rate and using less than half of the average CPU allocation. As shown in \Cref{tab:rq3_summary}, MLP and Poly $d{=}2$ reach similar or slightly higher success but consume more energy, while Linear, Poly $d{=}3$, and Bayesian regression lose substantial success. XGBoost therefore provides the strongest energy-oriented operating point.

For OCR, the static baseline is already close to optimal because PaddleOCRv4 with $1.5$ CPUs reaches $90.2\,\%$ success at $21.8\,\mathrm{kJ}$. \textsc{PRISM}+XGBoost reaches $91.0\,\%$ success at $21.5\,\mathrm{kJ}$ with only $0.6$ CPUs on average. The gain in energy is therefore small, but the reduction in average CPU allocation is substantial. The OCR plot in \Cref{fig:alpr-eval} also shows why the dynamic gains are smaller than for DETECT. Many fixed OCR configurations already operate in a narrow energy range, while DETECT has larger differences between lightweight and heavy detector configurations.

\textbf{Interpretation.} \Cref{fig:alpr-eval} and \Cref{tab:rq3_summary} show that dynamic selection is most beneficial when the static design space contains both cheap but infeasible and expensive but reliable configurations. This is pronounced for DETECT and weaker for OCR. \textsc{PRISM}+XGBoost matches or slightly exceeds the best static success rates while using substantially lower average CPU allocation in both stages. The DETECT--OCR crop dependency is handled through per-request prediction, allowing \textsc{PRISM} to adapt to actual or estimated crop dimensions instead of relying on a fixed configuration. The results show that stage-local adaptation composes effectively within a multi-stage chain without requiring cross-stage redistribution.

\section{Discussion and Future Work}\label{sec:Discussion}

\textbf{Scope of the Contribution.}
\textsc{PRISM} is a stage-level adaptation mechanism for edge AI microservices. Its contribution is the integration of container-level energy measurement, prediction-guided model selection, and in-place CPU scaling into a deployable runtime loop. This scope is complementary to chain-level orchestration frameworks such as \textsc{MARQ}. While \textsc{MARQ} selects among alternative service chains and execution paths under application-level constraints, \textsc{PRISM} refines the execution of an individual stage by selecting the model variant and CPU allocation for each request. A chain-level orchestrator can therefore assign per-stage deadlines, quality thresholds, and CPU budgets, while \textsc{PRISM} realizes these bounds locally through predictive in-place adaptation. \textsc{PRISM} does not solve global pipeline scheduling, cross-stage deadline redistribution, or resource allocation across competing pipelines.

\textbf{Quality and Energy Scope.}
\textsc{PRISM} uses offline per-model \gls{qor} scores as runtime constraints. Configurations are feasible only if the model's reference quality satisfies the requested threshold. This enables reproducible selection without ground-truth labels at runtime, but it does not guarantee the quality of each individual request. Input difficulty, image content, and distribution shift are therefore not directly captured by the current offline \gls{qor} abstraction. Future work should combine offline scores with calibrated online signals such as confidence, temporal consistency, or input-complexity estimates. Similarly, \textsc{PRISM} optimizes CPU-package energy because CPU allocation is the actuation mechanism studied here. Full-system components such as DRAM, storage, networking, idle baselines, and model-loading costs remain relevant for full-system accounting, but are not directly controlled by the local CPU-scaling decisions and require platform-specific measurement models.

\textbf{Runtime Overhead and Model Preloading.}
The XGBoost-Timefold configuration achieves the best prediction quality but incurs non-negligible overhead, although it remains below the corresponding inference latency in our experiments. Because the candidate set is bounded by the deployed model variants and CPU options, direct enumeration would be equivalent in the evaluated deployment and is the preferred implementation for latency-critical settings. \textsc{PRISM} also assumes that candidate models are preloaded and kept in memory. This avoids model-loading delays during request processing, but increases the persistent memory footprint. Our cold-start results indicate that this trade-off is stage-dependent, as preloading is more useful for stages with costly model initialization than for lightweight stages. Future work should quantify this cost and integrate model loading, eviction, and placement into the orchestration layer.

\textbf{Generalization and Future Work.}
The evaluation uses \gls{ALPR} as a running example because it exposes the core pattern targeted by \textsc{PRISM}. A detection stage produces input-dependent downstream work, alternative model variants provide different quality and cost profiles, and deadline constraints make static overprovisioning inefficient. This pattern appears in many edge AI pipelines, including video analytics, industrial inspection, autonomous perception, and medical imaging. \textsc{PRISM} is therefore not tied to license plate recognition, but to pipelines with measurable pre-execution features, bounded configuration choices, and stage-local quality and deadline constraints. Chain-level orchestration with frameworks such as \textsc{MARQ} can complement this mechanism by assigning application-level paths and per-stage bounds. To our knowledge, no drop-in adaptive baseline currently combines Kubernetes-based microservice deployment, in-place CPU scaling, per-request model selection, CPU-package energy optimization, and offline \gls{qor} constraints. Future work should therefore evaluate this stage-local mechanism in additional domains and study its integration with chain-level orchestration and heterogeneous accelerators.

\section{Related Work}
\label{sec:RelatedWork}
Research on energy- and latency-aware AI inference broadly falls into two areas. The first estimates energy or runtime before execution. The second adapts serving systems at runtime through scaling, scheduling, or approximate model selection.

In the area of predictive methods, Getzner et al.~\cite{getzner_accuracy_2023} propose layer-wise regression to predict the energy consumption of deep neural networks by composing per-layer costs into architecture-level estimates, thus avoiding full execution on the target device. Their focus is model-level energy prediction rather than runtime integration in containerized microservice pipelines. Inference-serving systems address a related problem. Clipper~\cite{crankshaw_clipper_2017} provides a general-purpose low-latency prediction serving system with caching, batching, and adaptive model selection. INFaaS~\cite{romero_infaas_2021} automates model selection and resource management based on application-level latency and accuracy requirements. Nexus~\cite{shen_nexus_2019} focuses on efficient GPU-cluster serving for DNN applications, while Clockwork~\cite{gujarati_serving_2020} exploits predictable DNN inference times to meet tight request-level SLOs. These systems establish important mechanisms for model serving, scheduling, and SLO-aware adaptation. \textsc{PRISM} differs by focusing on containerized edge microservices, CPU-package energy measurements, and in-place vertical CPU scaling under stage-local deadline and offline \gls{qor} constraints.

In the area of self-adaptive systems, Razavi et al.~\cite{razavi_tale_2024} study coordinated horizontal and vertical scaling for DL inference pipelines and formulate the end-to-end SLO-driven allocation problem. In a complementary line, Razavi et al.~\cite{razavi_sponge_2024} present \emph{Sponge}, which tackles per-request dynamic SLOs by combining in-place vertical scaling, dynamic batching, and request reordering, explicitly accounting for bandwidth-induced variability in mobile networks. Rattihalli et al.~\cite{rattihalli_exploring_2019} explore non-disruptive vertical autoscaling in Kubernetes, while Kaushik et al.~\cite{kaushik_study_2022} study power-aware vertical scaling for deadline-constrained containers. At the model-selection level, Matathammal et al.~\cite{matathammal_edgemlbalancer_2025} introduce \emph{EdgeMLBalancer}, an $\varepsilon$-greedy strategy that switches between object detectors to balance runtime and accuracy under resource constraints. Nigade et al.~\cite{nigade_jellyfish_2022} guarantee timely serving in dynamic wireless settings with \emph{Jellyfish} by jointly adapting DNNs and inputs to satisfy end-to-end latency SLOs while maintaining accuracy. Reinforcement-learning approaches explore the joint optimization space. Mounesan et al.~\cite{mounesan_infer-edge_2025} optimize edge inference parameters, including variant choice and partitioning/offloading decisions, to trade off latency, accuracy, and energy.

Hardware-centric approaches adjust low-level knobs to co-optimize efficiency. EcoEdgeInfer~\cite{rachuri_ecoedgeinfer_2024} shows that tuning CPU/GPU frequencies and batching can substantially change latency–energy trade-offs on Jetson-class edge devices and uses dynamic configuration to minimize violations under varying conditions. Zhang et al.~\cite{zhang_e4_2025} integrate early exits with DVFS in \emph{E4} to save energy for edge video analytics, adapting exit points and frequencies per frame complexity. Beyond serving stacks, AxIS~\cite{ghosh_approximate_2020} demonstrates that cross-layer approximations, spanning sensing, networking, storage, and inference, yield significant efficiency gains when quality loss is bounded. For serverless workflows, Rastegar et al.~\cite{rastegar_enex_2024} formulate energy-aware scheduling of function chains with fixed deadlines (EneX) via optimization and online scheduling to minimize active energy while avoiding cold starts.

Compared to these lines of work, \textsc{PRISM} targets stage-local adaptation in containerized edge microservice chains. It combines lightweight prediction of execution time and CPU-package energy from pre-execution features with per-request model selection and in-place vertical CPU scaling. This closes the loop from container-level measurement to runtime enforcement. Unlike global serving systems, chain-level schedulers, or vertical scalers for black-box applications, \textsc{PRISM} jointly selects the model variant and CPU allocation of each stage under deadline, energy, and offline \gls{qor} constraints. It does not redistribute budgets across stages or optimize cluster-wide resource allocation, but provides a deployable local control mechanism that can complement higher-level orchestrators.

\section{Conclusion}\label{sec:conclusion}

This paper introduced \textsc{PRISM}, a prediction-guided framework that unifies container-level monitoring, regression-based prediction, in-place vertical scaling, and per-stage model selection for microservice chains at the edge. \textsc{PRISM} selects CPU allocations and model variants under energy--time--quality constraints, reducing CPU-package energy for the detection stage by over one-third while maintaining a comparable success rate with less than half the average CPU allocation. In an evaluation with more than $52{,}000$ requests across both stages of an \gls{ALPR} pipeline, \textsc{PRISM} adapts models and CPU allocations per request, outperforming strong static baselines for detection and matching near-static-best performance for recognition with lower average CPU allocation. The approach is practical, directly deployable on Kubernetes, and grounded in container-level measurements, making it reproducible for edge microservice deployments. \textsc{PRISM} shows that combining accurate prediction with in-place scaling and per-stage model selection is an effective mechanism for reliable, energy-efficient execution of latency-sensitive AI services at the edge.
\section{Acknowledgment}
This work has been funded by the European Commission Horizon Europe Smart Networks and Services Joint Undertaking (SNS JU) EXIGENCE Project (Grant Agreement No. 101139120).

%----------------------------------------------------------------------------------------
%	BIBLIOGRAPHY
%----------------------------------------------------------------------------------------
\clearpage
\balance

\renewcommand*{\bibfont}{\footnotesize}
\printbibliography

%----------------------------------------------------------------------------------------
 
\end{document}